%% file: IEEE-conference-template-062824.tex
\documentclass[conference]{IEEEtran}
\IEEEoverridecommandlockouts
\usepackage{booktabs}
\usepackage{amssymb}
\usepackage{pifont}
\usepackage{multirow}
\usepackage{cite}
\usepackage{amsmath,amssymb,amsfonts}
\usepackage{algorithmic}
\usepackage{graphicx}
\usepackage{textcomp}
\usepackage{xcolor}
\usepackage{url}
\usepackage{hyperref}
\usepackage{cuted}
\usepackage{float}
\usepackage{siunitx}

\usepackage{algorithm,algorithmic}
\def\BibTeX{{\rm B\kern-.05em{\sc i\kern-.025em b}\kern-.08em
    T\kern-.1667em\lower.7ex\hbox{E}\kern-.125emX}}
\begin{document}

\title{Contrastive Multi-Task Learning with Solvent-Aware Augmentation for Drug Discovery
}

\author{
\IEEEauthorblockN{
Jing Lan\textsuperscript{*1}\thanks{*Co-first author},
Hexiao Ding\textsuperscript{*1},
Hongzhao Chen\textsuperscript{1},
Yufeng Jiang\textsuperscript{1},
Nga-Chun Ng\textsuperscript{1,2}\\
Gerald W.Y. Cheng\textsuperscript{1},
Zongxi Li\textsuperscript{3},
Jing Cai\textsuperscript{1},
Liang-ting Lin\textsuperscript{1},
Jung Sun Yoo\textsuperscript{\#1}\thanks{\#Corresponding author}
}
\IEEEauthorblockA{
\textsuperscript{1}Department of Health Technology and Informatics, The Hong Kong Polytechnic University, Hong Kong SAR, China\\
\textsuperscript{2}Department of Nuclear Medicine and PET, Hong Kong Sanatorium and Hospital, Hong Kong SAR, China\\
\textsuperscript{3}School of Data Science, Lingnan University, Hong Kong SAR, China\\
Emails: \{jing-hti.lan, hexiao.ding, hongzhao.chen, yufeng.jiang\}@connect.polyu.hk\\
\{wai-yeung.cheng, jing.cai, ltlin, jungsun.yoo\}@polyu.edu.hk\\
\{sam.nc.ng\}@hksh.com, \{zongxili\}@ln.edu.hk\\
}
}



\maketitle

\begin{abstract}
Accurate prediction of protein–ligand interactions is essential for computer-aided drug discovery. However, existing methods often fail to capture solvent-dependent conformational changes and lack the ability to jointly learn multiple related tasks. To address these limitations, we introduce a pre-training method that incorporates ligand conformational ensembles generated under diverse solvent conditions as augmented input. This design enables the model to learn both structural flexibility and environmental context in a unified manner. The training process integrates molecular reconstruction to capture local geometry, interatomic distance prediction to model spatial relationships, and contrastive learning to build solvent-invariant molecular representations. Together, these components lead to significant improvements, including a 3.7\% gain in binding affinity prediction, an 82\% success rate on the PoseBusters Astex docking benchmarks, and an area under the curve of 97.1\% in virtual screening. The framework supports solvent-aware, multi-task modeling and produces consistent results across benchmarks. A case study further demonstrates sub-angstrom docking accuracy with a root-mean-square deviation of 0.157 angstroms, offering atomic-level insight into binding mechanisms and advancing structure-based drug design.
Code is available at \url{https://github.com/1anj/SolvCLIP}.

\end{abstract}


\input{section/1_introduction}

\input{section/3_methods}
\input{section/4_experiments}
\input{section/5_results_discussion}
\input{section/6_conclusion}

\bibliography{ref}  
\bibliographystyle{IEEEtran}  
\input{section/7_appendix}
\end{document}

%% file: section/1_introduction.tex
\section{Introducion}
Accurate prediction of protein--ligand interactions is a fundamental objective in computer-aided drug discovery, particularly in virtual screening.  
The primary task is to determine whether a small molecule binds to a specific protein target, commonly referred to as the binary classification problem of ``bind or not''~\cite{ref1}. 
In addition to this, two other tasks are essential for evaluating molecular efficacy. 
One involves predicting the binding affinity, which is formulated as a regression problem~\cite{ref2}. The other focuses on estimating the docking pose of both the protein and the ligand, treated as a reconstruction problem~\cite{ref3}. 
Together, these tasks form the computational foundation for an effective virtual screening. Traditional molecular docking methods, which often rely on rigid-body approximations and empirical scoring functions, have provided valuable insights~\cite{ref4,ref5}. However, these approaches are frequently limited in their ability to model the full complexity of biochemical interactions, especially when faced with flexible ligands, induced fit effects, and solvent-dependent conformational dynamics~\cite{ref6,ref7}. 

In recent years, the integration of deep learning into molecular modeling has introduced transformative potential.  
The molecular-docking landscape has undergone a transformative computational revolution, epitomized by advanced deep learning models such as Uni-Mol~\cite{ref8}, EquiBind~\cite{ref9}, E3Bind~\cite{ref10}, TANKBind~\cite{ref11}, and KarmaDock~\cite{ref12}, which collectively challenge traditional interaction-prediction paradigms by introducing geometric learning and molecular representation strategies.  
Uni-Mol introduces a unified molecular representation framework that overcomes domain-specific limitations by using graph neural networks to capture intrinsic geometric and chemical features~\cite{ref8}.  
Building on this, EquiBind employs an equivariant neural architecture that rigorously enforces physical symmetries, resulting in more accurate binding-pose predictions that reflect the complex spatial nature of molecular interactions~\cite{ref9}.  
Further developments are exemplified by E3Bind and TANKBind, which push the boundaries of molecular modeling through integration with probabilistic reasoning and enhanced graph-based representations.  
E3Bind incorporates uncertainty quantification in its predictions, offering a transparent and interpretable approach to molecular interaction analysis~\cite{ref10}.  
In parallel, TANKBind excels in identifying intricate interaction patterns, particularly within challenging binding-site geometries~\cite{ref11}.  
Complementing these frameworks, KarmaDock combines geometric learning with sophisticated scoring strategies to improve binding-affinity predictions across a range of protein families~\cite{ref12}.  
Together, these methods represent an evolution in deep learning-based molecular representation and binding analysis, facilitating more precise and insightful modeling of molecular behavior.

Among existing methodologies, contrastive learning has shown considerable promise in modeling relational patterns between chemical and biological entities.  
Co-supervised Pre-training of Pocket and Ligand (CoSP) advances the field by applying contrastive learning to jointly embed pockets and ligands~\cite{ref13}.
In addition to this, DrugCLIP reformulates interaction prediction as a binary classification task by using matched and mismatched protein-ligand pairs to differentiate binders from non-binders~\cite{ref1}. 
However, despite its strengths, DrugCLIP and similar contrastive models often fall short in capturing nuanced biophysical interactions~\cite{ref14}.  
Furthermore, in pursuit of screening efficiency, DrugCLIP adopts a conservative model configuration that leads to a higher false negative rate~\cite{ref1}.  
Its single-task design further limits practical utility, as chemists must invest additional time in downstream affinity quantification and chemical-structure characterization.  
These constraints underscore the necessity for a more comprehensive, end-to-end solution that integrates multiple predictive tasks into a unified model.  
Such a framework would combine binding-pose estimation, affinity prediction, and interaction classification, thereby enabling faster and more accurate virtual screening while minimizing the need for manual post-processing.

Moreover, a critical limitation in most current predictive models lies in the inadequate consideration of solvent environments.  
Solvent effects play an important role in modulating molecular recognition, influencing both the energetic and geometric aspects of the binding process.  
Differences in solvent polarity, hydrogen-bonding capability, and dielectric properties can lead to substantial variations in binding poses and free-energy landscapes~\cite{ref15}.  
Specifically, a ligand may adopt a favorable conformation and exhibit strong binding affinity in an aqueous environment, but the same molecule might behave quite differently in a membrane-mimetic or organic solvent~\cite{ref15,ref16}.  
Neglecting such environmental variability reduces the ecological validity of predictions and impairs the model's ability to generalize across different biochemical conditions.

To address the limitations of conventional predictive models, we present a contrastive deep learning framework that incorporates solvent-specific enhancements to molecular conformations.  
By simulating environmental perturbations and embedding attention mechanisms, the model adapts both ligand and protein representations to varying solvent conditions.  
Pre-training involves molecular masking and reconstruction to refine feature learning and optimize performance across tasks such as binding classification, affinity estimation, and pose prediction.  
Solvent-aware data augmentation further strengthens the model's ability to capture context-dependent interactions and structural diversity.  
\textbf{This integrated approach advances structure-based drug design by delivering robust, accurate, and generalizable pre-training framework. As stated above, our main contributions are as follows:}
\begin{enumerate}
  \item We propose a solvent-aware molecular representation that captures conformational variability under diverse chemical environments.
  \item Contrastive learning combined with attention mechanisms enables adaptive modeling of protein–ligand interactions, leading to improvements in prediction accuracy.
  \item Geometry-enhanced self-supervised learning tasks that leverage global and local structural representation learning to enhance fine-tuned downstream task performance.
\end{enumerate}

%% file: section/3_methods.tex
\section{Methodology}
\begin{figure}[h]
    \includegraphics[width=\linewidth]{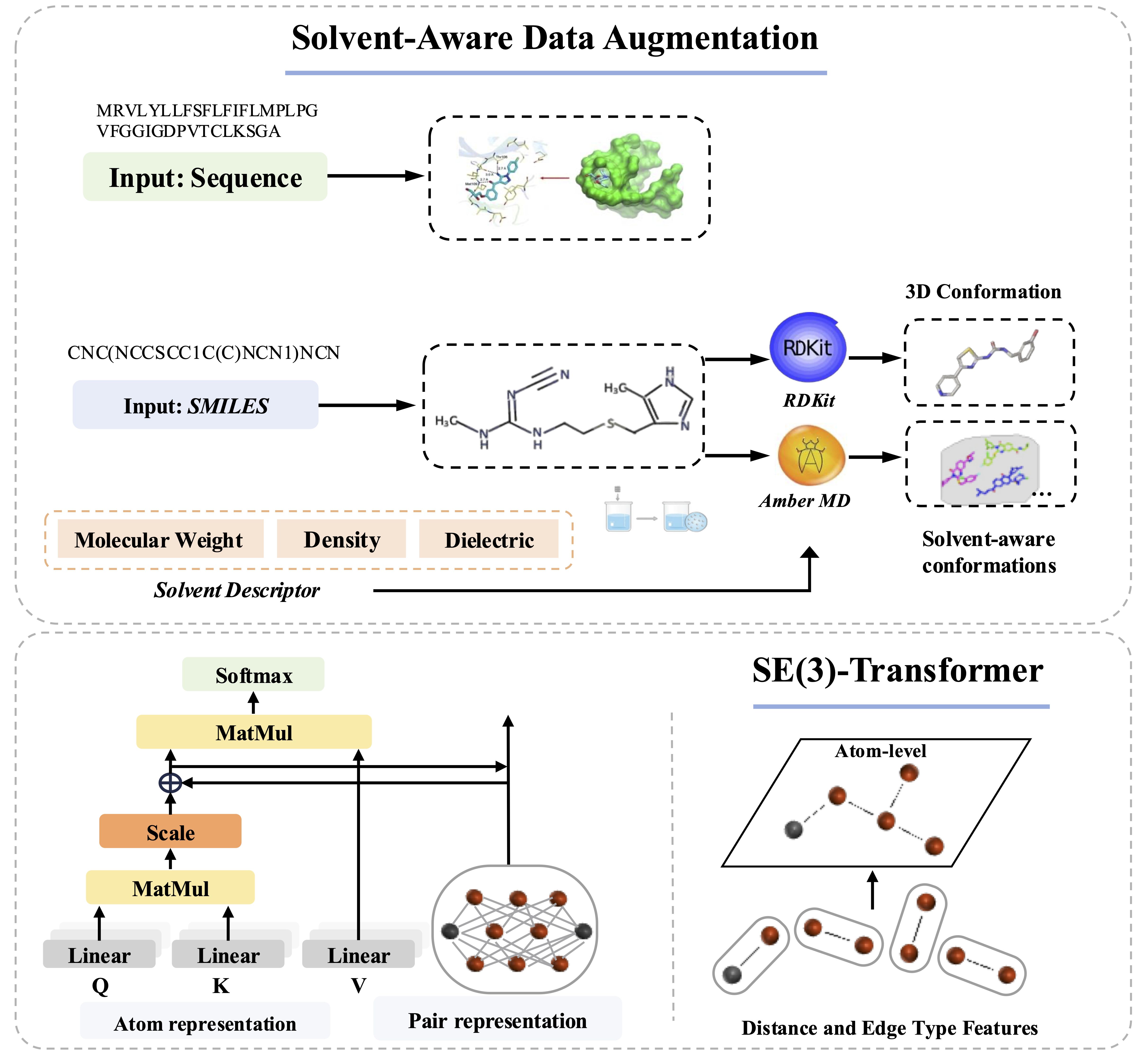}
    \caption{\textbf{Solvent-Aware Data Augmentation Pipeline.} \textbf{(a)} Protein sequences and ligand SMILES are transformed into solvent-aware 3D conformer ensembles using RDKit and AMBER-MD; \textbf{(b)} The SE(3)-Transformer encodes pocket-ligand graphs based on interatomic distances and edge types; \textbf{(c)} Contrastive learning is employed to align the pocket-ligand complex with its solvent-augmented counterpart.}
    \label{fig:solvent}
\end{figure}
\begin{figure*}[t]
  \centering
  \includegraphics[width=0.95\textwidth]{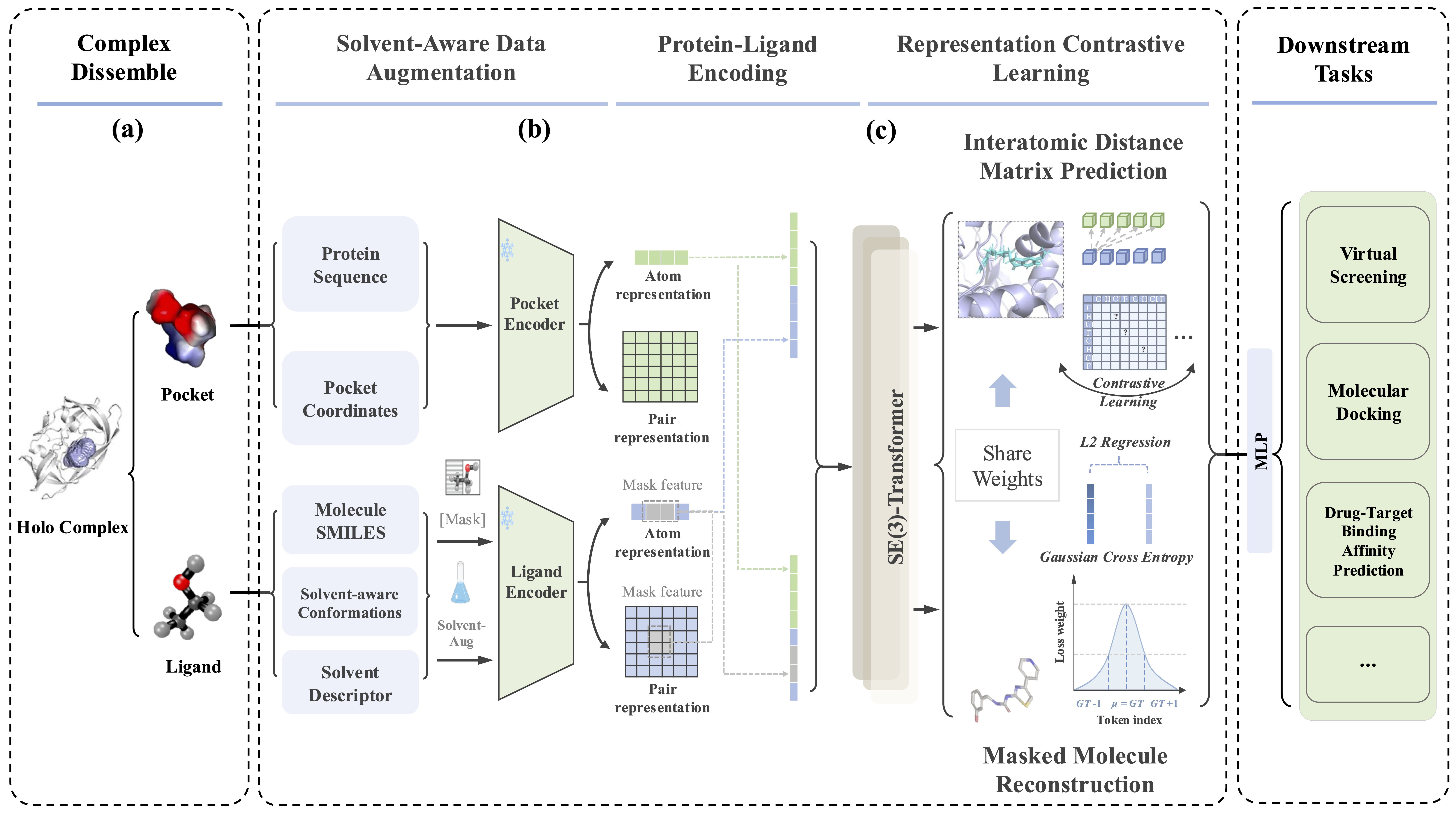}
  \caption{\textbf{Overview of the pre-training framework.} \ 
  \textbf{(a)} The complex is decomposed into the apo pocket and the ligand. The ligand is augmented with solvent-aware conformers;  
  \textbf{(b)} Two encoders are employed to generate initial pocket and ligand representations, with their parameters kept frozen throughout both training and inference;
  \textbf{(c)} A solvent-aware interaction module refines the embeddings for downstream tasks via contrastive self-supervised learning.}
  \label{fig:framework}
\end{figure*}

\subsection{Overview}
To capture intricate protein–ligand interactions influenced by solvent-dependent conformational diversity, we propose a novel pre-training framework specifically designed to model solvent effects. To evaluate its effectiveness, we design three complementary downstream experiments, including \textbf{\textit{ligand binding affinity prediction}}, \textbf{\textit{molecular docking}}, and \textbf{\textit{virtual screening}}. Within this framework, the objective for pretraining is the direct prediction of the protein–ligand complex structure, given the rigid apo pocket structure and an ensemble of ligand conformers. Namely, the provided protein structure preserves its conformation from the bound complex. This formulation captures the structural rigidity of the protein versus the inherent flexibility of the ligand, particularly its solvent-modulated torsional states, enhancing physiological relevance over single-conformation approaches~\cite{ref17}.

\subsubsection*{Graph Definitions}
Formally, the ligand is represented as an atom-level graph:
\begin{equation}
\mathcal{G}_{\text{L}}=(\mathcal{T}_{\text{L}}, \mathcal{C}_{\text{L}}),\qquad 
\mathcal{T}_{\text{L}}\in\mathbb{N}^{N_{\text{L}}},\; \mathcal{C}_{\text{L}}\in\mathbb{R}^{N_{\text{L}}\times 3}
\end{equation}
where $\mathbf{t}_i \in \mathcal{T}_\text{L}$ denotes the atom type and $\mathbf{c}_i \in \mathcal{C}_\text{L}$ represents 3-D coordinates of atom $i$.

Analogously, the protein pocket is defined as:
\begin{equation}
\mathcal{G}_{\text{P}}=(\mathcal{T}_{\text{P}}, \mathcal{C}_{\text{P}}),\qquad 
\mathcal{T}_{\text{P}}\in\mathbb{N}^{N_{\text{P}}},\; \mathcal{C}_{\text{P}}\in\mathbb{R}^{N_{\text{P}}\times 3}
\end{equation}

The protein-ligand complex is then formulated as graph:
\begin{equation}
\mathcal{G}_{Complex} = (\mathcal{T}_{Complex}, \mathcal{C}_{Complex})
\end{equation}

which is partitioned into pocket ($\mathcal{G}_\text{P} = (\mathcal{T}_\text{P}, \mathcal{C}_\text{P})$) and ligand ($\mathcal{G}_\text{L} = (\mathcal{T}_\text{L}, \mathcal{C}_\text{L})$) subgraphs.

\subsubsection*{Protein and Molecule Representation}
To achieve geometrically faithful representations of both ligands and protein binding pockets, we adopt the unified encoder backbone from DrugCLIP~\cite{ref1}, which integrates a pre-trained SE(3)-equivariant transformer architecture based on UniMol~\cite{ref8}.
The encoder initializes pairwise geometric features $q_{ij}^{(0)}$  and the self-attention mechanism integrates geometric information as spatial bias in each transformer layer $l \in \{1,\dots,L\}$:
\begin{equation}
\mathrm{Attention}^{(l)}\left(Q^l, K^l, V^l\right) = \mathrm{softmax}\left( \frac{Q^l (K^l)^\top}{\sqrt{d}} + q_{ij}^{l} \right) V^l
\end{equation}
with the pairwise representation updated recursively:
\begin{equation}
q_{ij}^{l+1} = q_{ij}^{l} + \frac{q_i^l (K_j^l)^\top}{\sqrt{d}}
\end{equation}
Through iterative refinement, this process propagates invariant geometric features across attention layers, facilitating 3-D representation learning while maintaining SE(3) equivariance.

\subsubsection*{Solvent-Aware Data Augmentation} 

For each ligand, to ensure the primary molecular structure is used, we first generate an initial 3-D structure with RDKit based on the chemical information of ligand. The atom and edge attributes were calculated using the RDKit libraries, and molecular graphs were generated with the PyTorch Geometric libraries.
If conformer generation fails, we fall back to the ETKDG conformer generator by default. Specifically, we perturb the torsion angles and atomic coordinates with Gaussian noise to obtain an unbiased, unbound-like geometry to approximate the unbound state. The strategy follows the formulation:
\begin{equation}
    \mathcal{G}_{\widehat{\text{L}}} = \bigl(\mathcal{T}_{\widehat{\text{L}}}, \mathcal{C}_{\widehat{\text{L}}}\bigr)
\end{equation}
where $\mathcal{T}_{\widehat{\text{L}}}$ denotes the perturbed torsional topology and $\mathcal{C}_{\widehat{\text{L}}}$ represents the corresponding atomic coordinates of the unbound-like ligand geometry.

Furthermore, we introduce solvent-aware data augmentation utilizing the GNN-based implicit solvent approach~\cite{ref15}, which emulates quantum mechanical solvent effects through a graph neural network trained on classical molecular dynamics (MD) data. This approach enables rapid generation of solvent-specific conformational ensembles with accuracy comparable to explicit-solvent MD simulations, while significantly reducing computational demands. For each ligand, we sample 39 conformers $\{\mathcal{G}^{\,s}_{\text{L}}\}_{s=1}^{39}$ representing \textit{Boltzmann-weighted distributions} across 39 organic solvents, thereby capturing both enthalpic and entropic contributions to molecular flexibility as illustrated in Fig.~\ref{fig:solvent}. The resulting solvent-aware ligand structure is defined as $\mathcal{G}^{\,s}_{\text{L}}=(\mathcal{T}^{\,s}_{\text{L}}, \mathcal{C}^{\,s}_{\text{L}})$.

\subsubsection*{Pre-training Framework}  
As illustrated in Fig.~\ref{fig:framework}, given a protein pocket $\mathcal{G}_{\text{P}}$ and a ligand $\mathcal{G}_{\text{L}}$, the preliminary representations are secured by pre-trained encoders: $\mathcal{E}_{\text{P}}$ for the pocket and $\mathcal{E}_{\text{L}}$ for the ligand. 
\begin{equation}
\mathcal{R}_{\text{P}}^{(0)} = \mathcal{E}_{\text{P}}(\mathcal{G}_{\text{P}}), \qquad
\mathcal{R}_{\text{L}}^{(0)} = \mathcal{E}_{\text{L}}(\mathcal{G}_{\text{L}}).
\end{equation}

Our framework provides architectural flexibility by accommodating diverse state-of-the-art pocket or ligand encoder architectures. Throughout both pre-training and fine-tuning phases, all operations are confined to the latent representation space, with the weights of $\mathcal{E}_{\text{P}}$ and $\mathcal{E}_{\text{L}}$ remaining frozen. This design explicitly highlights the learning capacity of the subsequent \emph{protein-ligand solvent-aware interaction} module.

The interaction module, implemented as an $N$-layer SE(3)-equivariant transformer $\theta_I$, processes atomic-level and pairwise representations to generate refined embeddings:
\begin{align}
\mathcal{R}_{\text{PL}}^{(N)} = \theta_I^{(N)}\left( \mathcal{R}_\text{P}^{(0)}, \mathcal{R}_\text{L}^{(0)} \right)
\end{align}
where $\mathcal{R}_{\text{PL}}^{(N)}$ denotes the pairwise binding representation capturing complex interaction patterns.

\subsection{Objectives for Self-supervised Learning}
\label{section:distance}
Classical biophysical studies demonstrate that protein-ligand binding affinities strongly correlate with intermolecular distances~\cite{ref21,ref22}, a principle extensively adopted in scoring functions~\cite{ref23} and deep learning frameworks~\cite{ref24,ref25}. Building on this foundation and motivated by the efficacy of masked reconstruction in BERT-based models for capturing complex interactions such as MoleBERT~\cite{ref20}), we propose two synergistic objectives for self-supervised learning. These objectives integrate geometry-enhanced molecular representations~\cite{ref26} through weight-sharing to enhance 3-D geometric reasoning: (1) a global geometric prediction task and (2) a local geometric reconstruction task, as detailed in Fig.~\ref{fig:framework}.

Among these, the global geometric task \textit{Interatomic Distance Matrix Prediction (IDMP)} frames atomic distance regression through pairwise distance matrix construction. This approach enables the model to learn interatomic spatial relationships, thereby acquiring a comprehensive understanding of molecular geometry and global spatial organization. We emphasize fine-grained interaction learning through primary ligand conformations. Specifically, the ligand $\mathcal{G}_{\widehat{\text{L}}}$ and pocket $\mathcal{G}_\text{P}$ serve as inputs to predict the protein-ligand complex's pairwise distance matrix $\mathbf{\mathcal{D}}$, where elements $d_{ij} \in \mathbf{\mathcal{D}}$ denote Euclidean distances between pocket atom $i$ and ligand atom $j$. This objective function is formalized as:
\begin{equation}
\mathcal{L}_{reg} = \frac{1}{|\Omega|} \sum_{(i,j) \in \Omega} ( \hat{d}_{ij} - d_{ij} )^2
\end{equation}
where $\mathcal{L}_{reg}$ denotes the \textit{$L_2$ regression loss}, $\Omega$ is the set of protein-ligand atom pairs, $\hat{d}_{ij}$ is the predicted distance, and $d_{ij}$ is the ground truth distance.

The local geometric reconstruction task \textit{Masked Molecular Reconstruction (MMR)} models conditional dependency relations between protein and ligand representations during binding. Concretely, we replace atom-level embeddings in both the primary ligand conformation ($\mathcal{G}_{\widehat{\text{L}}}$) and solvent-modulated conformers ($\mathcal{G}^{s}_{\text{L}}$) with learnable masked placeholders $\mathbf{q}_{\text{mask}} \in \mathbb{R}^{d}$, and randomly masking a subset of pocket atom tokens $\mathcal{G}_{\text{P}}$. To enable continuous value regression, we introduce \textit{Gaussian Cross-Entropy (GCE) Loss}~\cite{ref18}, softening standard cross-entropy with a Gaussian kernel over the atom-type vocabulary.
\begin{equation}
\mathcal{L}_{GCE} = -\sum_{i \in \mathcal{M}} \sum_{c=1}^{C} \mathcal{N}(\textbf{c}|\textbf{c}_i, \sigma) \log P(\hat{\textbf{c}}_i = c)
\end{equation}
where $\mathcal{M}$ denotes masked positions, $\textbf{c}_i$ is the ground truth atom type, $\hat{\textbf{c}}_i$ is the predicted type, $C$ is the atom-type vocabulary size, and $\mathcal{N}$ is a Gaussian kernel that incorporates chemical similarity to soften standard cross-entropy.

The reconstruction target is defined by:
\begin{equation}
\mathcal{L}_{Recon} = (\mathcal{L}_{reg}^{\mathcal{G}_{\widehat{\text{L}}}}
+ \sum_{s=1}^{39}\mathcal{L}_{reg}^{\mathcal{G}^{s}_{\text{L}}}) + \mathcal{L}_{GCE}
\end{equation}
where $\mathcal{L}_{reg}$ denotes the \textit{$L_2$ loss} for reconstructing the target embeddings, and $\mathcal{L}_{\text{GCE}}$ supervises the masked pocket tokens.

\subsection{Objectives for Contrastive Learning}
DrugCLIP~\cite{ref1} reformulates virtual screening as a similarity-matching problem, focusing on binding likelihood rather than binding affinity or pose specifics. We extend this paradigm by expanding the similarity metric from \emph{pocket-ligand} to \emph{solvent-aware pocket-ligand complexes}.

To further enhance representation robustness across solvent environments, we introduce a \textit{Contrastive Learning (CL)} algorithm inspired by SimCLR~\cite{ref19}.  This component enforces geometric consistency across solvent-modulated binding states while distinguishing functionally irrelevant variations.

\begin{figure}[htpb]
    \centering
    \includegraphics[width=0.8\linewidth]{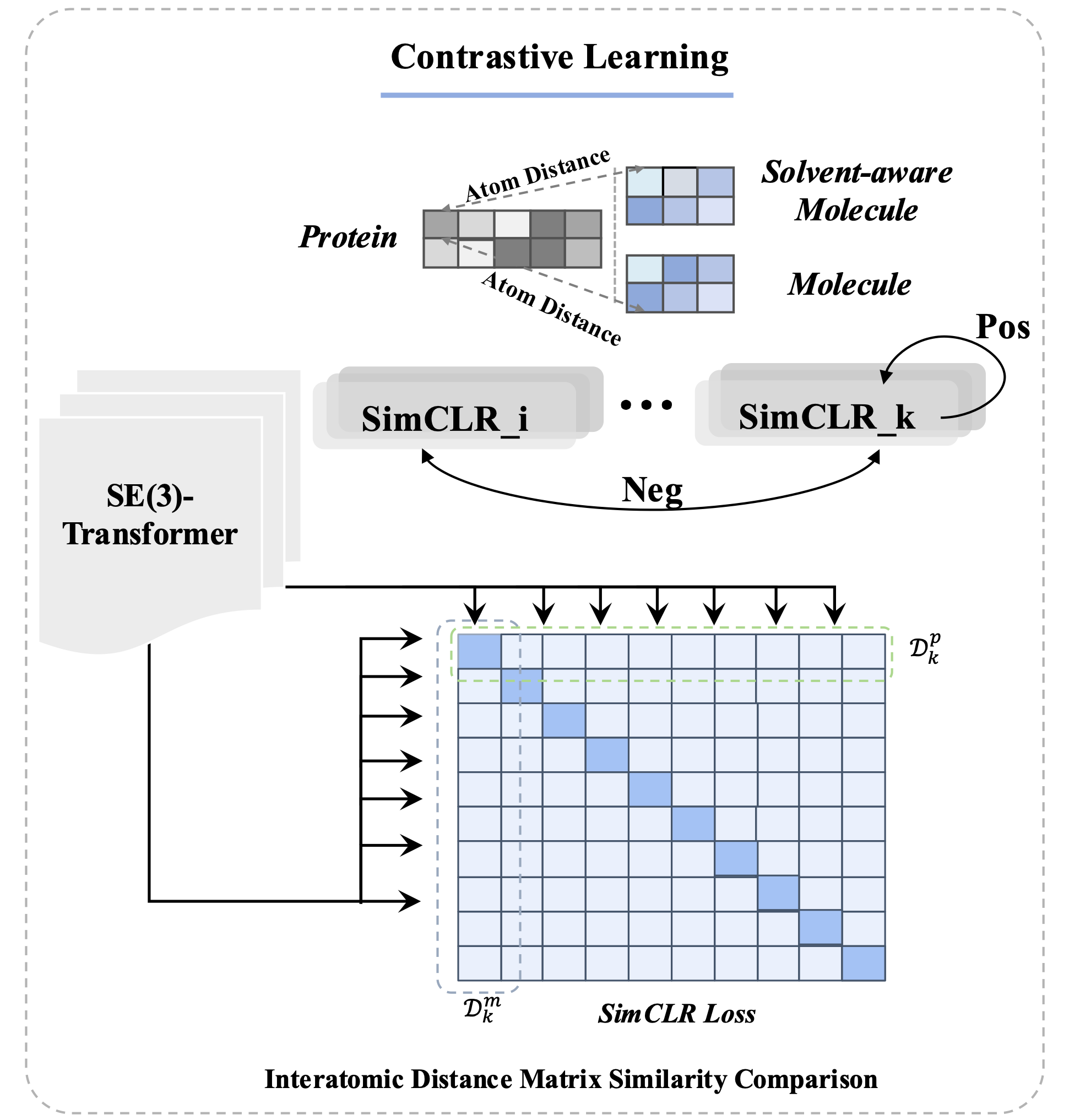}
    \caption{\textbf{Contrastive Learning (CL).} Align interatomic distance matrix of the same pocket-ligand pair (solvent vs. non-solvent) as positives; negatives swap ligand or pocket within the batch.}
    \label{fig:loss}
\end{figure}

As illustrated in Fig.~\ref{fig:loss}, the contrastive learning algorithm operates on distance matrix representations derived from different ligand states. Positive pairs align distance matrices of identical pocket-ligand complexes under distinct solvent conditions $(\mathcal{D}_{\widehat{L}|P}, \mathcal{D}_{L^s|P})$. Negative pairs are formed by substituting ligand or pocket from different complexes within the batch $(\mathcal{D}_{\widehat{L}|P}, \mathcal{D}_{L^*|P^*})$ where $P^* \neq P$ or $L^* \neq L$.
The contrastive loss maximizes similarity for positive pairs while minimizing similarity for negative pairs:
\begin{equation}
\mathcal{L}_{\text{cont}} = -\log \frac{
    \exp \left( \operatorname{sim}(\mathcal{D}_{\widehat{L}|P}, \mathcal{D}_{L^s|P}) / \tau \right)
}{
    \sum\limits_{(\mathcal{D}_{\widehat{L}|P}, \mathcal{D}_{L^*|P^*}) \in \mathcal{N}} \exp \left( \operatorname{sim}(\mathcal{D}_{\widehat{L}|P}, \mathcal{D}_{L^*|P^*}) / \tau \right)
}
\end{equation}
where $\operatorname{sim}(\cdot,\cdot)$ denotes cosine similarity, $\tau$ is the temperature parameter, and $\mathcal{N}$ represents the negative pair set. This objective for pre-training integrates global geometry regression with solvent-adaptive discriminative capabilities.

%% file: section/4_experiments.tex
\section{Experiments}

\subsection{Training Data}
\subsubsection{Dataset Construction}
We curated 348,970 protein-ligand complexes from the January 2025 BioLip~\cite{ref27} release through a multi-stage filtration process: retaining only drug-like ligands bound to protein binding pockets; defining binding pockets as residues within an 8 Å radius of any ligand heavy atom; removing non-protein components (nucleic acids, linkers, crystallographic additives, and ions); and excluding 24,809 entries due to PDBbind-CN licensing restrictions.

\subsubsection{Solvent-Aware Conformer Generation}
To comprehensively model solvent-dependent conformational diversity, we generated ligand conformers in 38 distinct solvents spanning the full polarity continuum (ranked by descending static dielectric constant). This yielded 16,037,848 conformations through explicit-solvent molecular dynamics equilibration and clustering, with solvent-specific distributions detailed in Table~\ref{tab:solvent}. Following de-duplication and merging procedures, the final pre-training dataset comprises 311,123 solvent-augmented protein-ligand complexes.

\begin{table}[h]
\centering
\caption{Ligand conformations generated in different solvents}
\resizebox{\columnwidth}{!}{%
\begin{tabular}{lS[table-format=7.0]|lS[table-format=7.0]}
\hline
\textbf{Solvent} & {\textbf{Conformations}} & \textbf{Solvent} & {\textbf{Conformations}} \\
\hline
Methanol            & 1107512 & Hexafluoroacetone & 367490 \\
DMSO                & 1105447 & HMPA               & 367365 \\
Chloroform          & 1089696 & DCM                & 367312 \\
Butylformate        & 370327  & Benzonitrile       & 367008 \\
NMP                 & 370315  & Hexafluorobenzene  & 366831 \\
Nitromethane        & 370304  & Propionitrile      & 366828 \\
Sulfolane           & 370280  & Pyridine           & 366067 \\
IPA                 & 370158  & Carbon tetrachloride & 365678 \\
DMPU                & 369745  & Acetonitrile       & 365464 \\
Octanol             & 369453  & Cyclohexane        & 365364 \\
Ethyl acetate       & 369272  & Benzene            & 364541 \\
Glycerin            & 368570  & 1,4-Dioxane        & 364334 \\
DMF                 & 367528  & Acetone            & 364183 \\
Diethyl ether       & 364052 & 2-Nitropropane & 359196 \\
Nitrobenzene        & 363863 & MTBE           & 359136 \\
Trifluorotoluene    & 363727 & Acetic acid    & 359122 \\
THF                 & 361535 & Oxylol         & 358606 \\
Toluene             & 361517 & DME            & 308870 \\
Hexane              & 361073 & Ethanol        & 360079 \\
\hline
\end{tabular}%
}
\label{tab:solvent}
\end{table}
\subsection{Implementation Details}

The framework was implemented in PyTorch 2.7.1 and trained on 4 NVIDIA A800 GPUs. Optimization employed the Adam algorithm with a learning rate of 1e-4 and weight decay of 1e-4. During pre-training, the masked molecular reconstruction (masking probability = 0.15, masking ratio = 0.8, noise ratio = 0.8) and interatomic distance matrix prediction objectives were equally weighted. Training proceeded for 20 epochs with a global batch size of 32, using gradient clipping (norm = 1.0) to stabilize optimization.

%% file: section/5_results_discussion.tex
\section{Results \& Discussion}

\subsection{Performance Comparison in Downstream Tasks}

\subsubsection{Downstream Task 1. Ligand Binding Affinity}

\paragraph{Experimental Setup}
We evaluate our model primarily on the Ligand Binding Affinity (LBA) prediction task using the LBA dataset from the Atom3D benchmark~\cite{ref28}, containing protein-ligand complexes sourced from the PDBBind database~\cite{ref29}, partitioned into training (3,507 complexes), validation (466), and test (490) sets. Performance was assessed via Root Mean Square Error (RMSE), Pearson correlation, and Spearman correlation.

\paragraph{Baselines}
We benchmark our model against established families of affinity predictors. Sequence-based methods including DeepDTA~\cite{ref40}, TAPE, and ProtTrans utilize primary structural information. Structure-aware geometric networks such as SchNet, Equiformer, and ProNet~\cite{ref30} directly incorporate atomic coordinates, with ProNet representing the current state-of-the-art among fully-supervised baselines. Finally, pre-training and self-supervised approaches including SE(3)-DDM, Uni-Mol~\cite{ref8}, SMT-DTA, and GeoSSL  leverage unlabeled data to compensate for affinity labels.

\paragraph{Results and Analysis}
As demonstrated in Table~\ref{tab:lba}, the comparative results reveal consistent advantages of our approach across all evaluation dimensions. First, our method reduces RMSE by 7.7\% compared to the best baseline. Second, we outperform specialized geometric frameworks like ProNet~\cite{ref30} despite their task-specific architectures, demonstrating superior comprehensive interaction modeling. Third, compared to Uni-Mol~\cite{ref8}, we achieve an 8.7\% RMSE reduction. These results collectively demonstrate our framework's superior capacity for modeling solvent-modulated conformational dynamics in binding interactions.

\begin{table}[t]
\centering
\caption{Performance comparison on LBA dataset. 
Best values are in \textbf{bold}; second-best are \underline{underlined}.}
\resizebox{\columnwidth}{!}{%
\begin{tabular}{lccc}
\toprule
\textbf{Method} & \textbf{RMSE ↓} & \textbf{Pearson ↑} & \textbf{Spearman ↑} \\
\midrule
\multicolumn{4}{l}{\textit{Sequence-based deep learning}}\\
DeepDTA & 1.762	& 0.666 & 0.663 \\
TAPE & 1.633 & 0.568 & 0.571 \\
ProtTrans & 1.641 & 0.595 & 0.588 \\
\midrule
\multicolumn{4}{l}{\textit{Structure-based deep learning}}\\
SchNet & 1.406 & 0.565 & 0.549 \\
EGNN & 1.409 & 0.566 & 0.548 \\
ET & 1.367 & 0.599 & 0.584 \\
GemNet & 1.393 & 0.569 & 0.553 \\
MACE & 1.385 & 0.599 & 0.580 \\
Equiformer & 1.350 & 0.604 & 0.591 \\
LEFTNet & 1.377 & 0.588 & 0.576 \\
Atom3D-CNN & 1.621 & 0.608 &	0.615 \\
Atom3D-ENN & 1.620 & 0.623 & 0.633 \\
Atom3D-GNN & 1.408 & 0.743 & 0.743 \\
Holoprot & 1.365 & 0.749 & 0.742 \\
ProNet & \underline{1.343} & \underline{0.765} & \underline{0.761} \\
\midrule
\multicolumn{4}{l}{\textit{Pre-training / Self-supervised methods}}\\
SE(3)-DDM  &  1.451 &  0.577 &  0.572 \\
DeepAffinity & 1.893 & 0.415 & 0.426 \\
SMT-DTA & 1.347 & 0.758 &	0.754 \\
GeoSSL & 1.451 & 0.577 & 0.572 \\
Uni-Mol & 1.357	& 0.753	 & 0.750 \\
\midrule
\textbf{Ours} & \textbf{1.239} & \textbf{0.793} & \textbf{0.791} \\
\bottomrule
\end{tabular}
}
\label{tab:lba}
\end{table}

\subsubsection{Downstream Task 2. Molecular Docking}
\paragraph{Experimental Setup}
Adopting established targeted docking protocols, we utilized PDBbind v2020~\cite{ref29} for training and CASF-2016 as an independent test set containing 285 high-quality protein-ligand complexes. To ensure data integrity, we implemented Uni-Mol~\cite{ref8}'s filtering protocol, excluding training complexes that exhibit significant similarity to test set proteins or ligands, yielding a refined training corpus of 18,404 ground-truth complexes.

\paragraph{Baselines}
Our comparative analysis includes representative methods spanning classical and modern docking approaches. We evaluate AutoDock4~\cite{ref31}, AutoDock Vina~\cite{ref32}, Vinardo~\cite{ref33}, and Smina~\cite{ref34}, which represent physics-driven methods relying on force fields and conformational sampling. 

\begin{figure*}[t] \centering \includegraphics[width=0.95\linewidth]{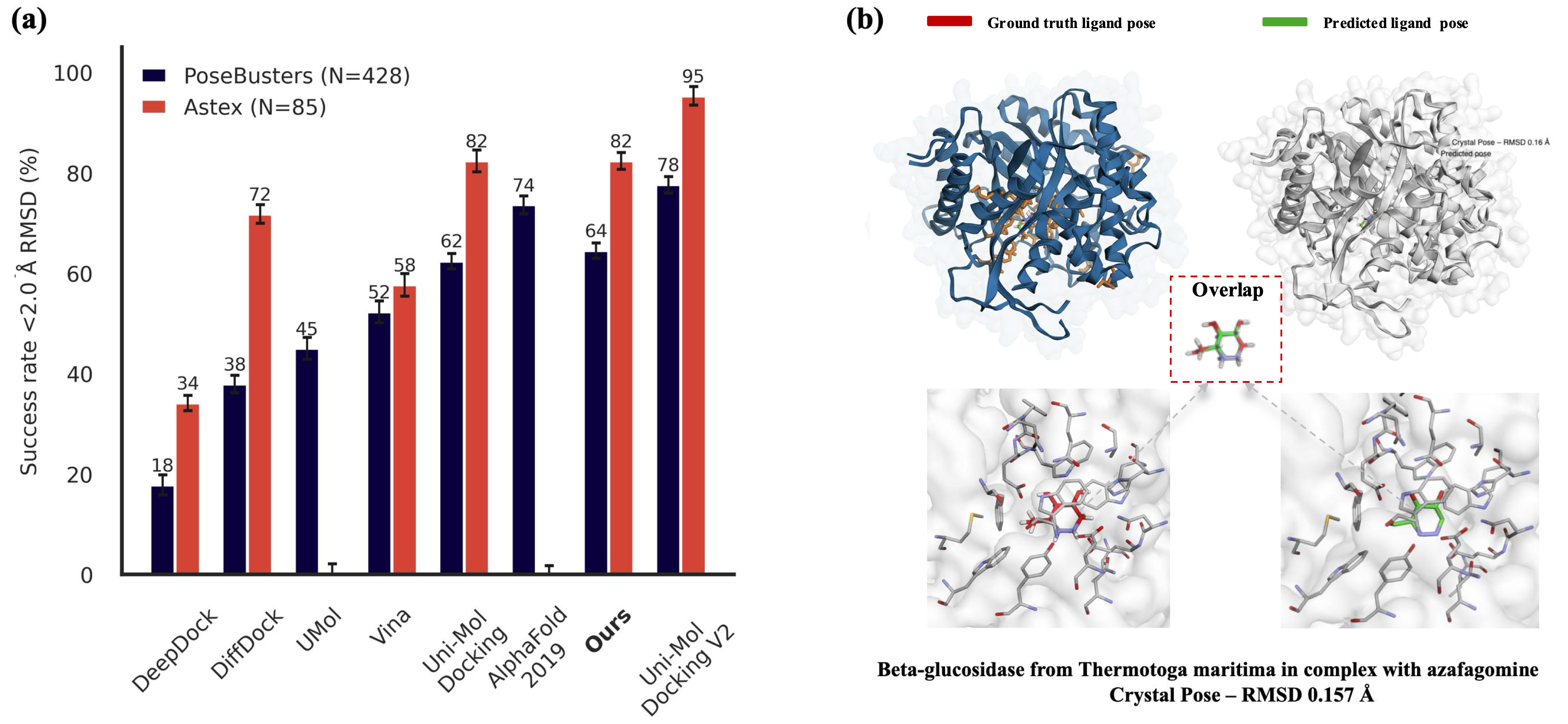} \caption{\textbf{Performance comparison on PoseBusters and Astex.} \textbf{(a)} overall success rates ($<$2.0 Å RMSD) across 428 PoseBusters complexes and 85 Astex entries for all evaluated methods; “\textbf{Ours}” is highlighted. \textbf{(b)} case-study of structural superposition for PDB entry \textbf{2J7H} (Thermotoga maritima $\beta$-glucosidase bound to azafagomine), showing the predicted pose (\textit{green}) overlaid on the crystallographic pose (\textit{red}).}\label{fig:pb} \end{figure*}

\paragraph{Results and Analysis}
Our model achieves the highest success rate among all evaluated methods when the RMSD threshold is set at $<$ 5.0 \AA\ in CASF-2016 (Table~\ref{tab:casf}).  In the more stringent PoseBusters (428 complexes) and Astex (85 complexes) benchmarks~\cite{ref35}, it delivers 64\% and 82\% success rates, respectively, at the $<$ 2.0 \AA\ cutoff, shown in Fig.~\ref{fig:pb}. Notably, since Uni-Mol Docking v2 was trained on a large amount of data with a training time three times longer than ours, these results demonstrate that our model can achieve competitive performance even after fine-tuning in a short period.

\begin{table}[h]
\centering
\caption{Performance comparison on protein-ligand binding pose prediction, following the results reported in Uni-Mol.\ Best values are in \textbf{bold}; second-best are \underline{underlined}.}
\resizebox{\columnwidth}{!}{%
\begin{tabular}{lccccr}
\toprule
\textbf{Method} & \textbf{\%$<$\,1.0\,\AA ↑} & \textbf{\%$<$\,2.0\,\AA ↑} & \textbf{\%$<$\,3.0\,\AA ↑} & \textbf{\%$<$\,5.0\,\AA ↑} & \textbf{Avg.\ (Å) ↓}\\
\midrule
AutoDock4 & 21.8 & 35.4 & 47.0 & 64.6 & 3.53\\
AutoDock Vina & \underline{44.2} & 64.6 & 73.7 & 84.6 & 2.37\\
Vinardo & 41.8 & 62.8 & 69.8 & 76.8 & 2.49\\
Smina & \textbf{47.4} & 65.3 & 74.4 & 82.1 & 1.84\\
Uni-Mol Docking & 43.2 & \textbf{80.4} & \textbf{87.0} & \underline{94.0} & \textbf{1.62}\\
\midrule
\textbf{Ours} & 41.1 & \underline{76.5} & \underline{86.3} & \textbf{94.4} & \underline{1.73} \\
\bottomrule
\end{tabular}
}
\label{tab:casf}
\end{table}

\subsubsection{Downstream Task 3. Virtual Screening}
\paragraph{Experimental Setup}
We benchmark our work on the DUD-E dataset~\cite{ref36}, which aggregates 102 pharmaceutically relevant targets spanning multiple protein families and, for each target, supplies an average of 224 active compounds alongside a curated set of more than 10,000 decoys that are physicochemically matched yet topologically distinct from the actives. Following the protocol established by AttentionDTI~\cite{ref37} and DrugVQA~\cite{ref38}, we implemented strict three-fold cross-validation to prevent data leakage. During fine-tuning, we froze the encoders and updated only the interaction module and classification head. Performance was evaluated using the Area Under the Curve (AUC) and Receiver Operating Characteristic (ROC) enrichment (RE) metrics.

\paragraph{Baselines}
Our study compares with classical docking (AutoDock Vina~\cite{ref32}); statistical learning (RF-score, NNScore); supervised deep learning (3D-CNN, Graph-CNN); and pre-training or self-supervised methods (CoSP~\cite{ref13}, Uni-Mol~\cite{ref8}, DrugVQA~\cite{ref38}, DrugCLIP\textsubscript{ZS}~\cite{ref1}). These represent the current state-of-the-art across methodological categories.

\paragraph{Results and Analysis}
Table~\ref{tab:dude} confirms our method's state-of-the-art performance in three RE metrics. First, near-identical AUC (97.1\% vs. DrugVQA's 97.2\%) demonstrates parity. Second, ROC enrichment at 5\% reveals solvent-aware modeling substantially boosts early enrichment at critical 0.5\%-2\% thresholds. Third, consistent improvements over DrugCLIP\textsubscript{ZS} validate geometric pre-training and task-specific fine-tuning, the reduced ROC enrichment further confirms enhanced stability from solvent-contextualized learning.

\begin{table}[t]
\centering
\caption{Performance comparison on DUD-E benchmark. 
Best values are in \textbf{bold}; second-best are \underline{underlined}.}
\resizebox{\columnwidth}{!}{%
\begin{tabular}{lccccc}
\toprule
\textbf{Method} & \textbf{AUC (\%) ↑} & \textbf{RE@0.5\% ↑} & \textbf{RE@1\% ↑} & \textbf{RE@2\% ↑} & \textbf{RE@5\% ↑} \\
\midrule
\multicolumn{6}{l}{\textit{Docking-based methods}}\\
AutoDock Vina & 71.6 & 9.14 & 7.32 & 5.81 & 4.44 \\
\midrule
\multicolumn{6}{l}{\textit{Scoring-function-based ML}}\\
RF-score & 62.2 & 5.63 & 4.27 & 3.50 & 2.68 \\
NNScore  & 58.4 & 4.17 & 2.98 & 2.46 & 1.89 \\
\midrule
\multicolumn{6}{l}{\textit{Supervised deep learning}}\\
3D-CNN    & 86.8 & 42.56 & 26.66 & 19.36 & 10.71 \\
Graph-CNN & 88.6 & 44.41 & 29.75 & 19.41 & 10.74 \\
\midrule
\multicolumn{6}{l}{\textit{Pre-training / Self-supervised methods}}\\
DrugVQA              & \textbf{97.2} & 88.17 & 58.71 & 35.06 & \textbf{17.39} \\
AttentionSiteDTI     & \underline{97.1} & \underline{101.74} & \underline{59.92} & \underline{35.07} & 16.74 \\
DrugCLIP\textsubscript{ZS} & 80.9 & 73.97 & 41.79 & 23.68 & 11.16 \\
CoSP                 & 90.1 & 51.05 & 35.98 & 23.68 & 12.21 \\
Uni-Mol              & 94.5 & 82.59 & 50.21 & 30.16 & 14.79 \\
\midrule
\textbf{Ours}  & \underline{97.1} & \textbf{109.20} & \textbf{61.67} & \textbf{36.61} & \underline{16.89} \\
\bottomrule
\end{tabular}
}
\label{tab:dude}
\end{table}

\subsection{Ablation Study}

We conducted a statistically rigorous ablation study using three-fold cross-validation on the DUD-E benchmark~\cite{ref36} to quantify the contributions of the components. Performance metrics (mean ± std) in Table~\ref{tab:ablation} establish the significance of core modules: \textit{MMR} for local geometry learning, \textit{IDMP} for global spatial modeling, and \textit{CL} for solvent-invariant representation learning.

\begin{table*}[t]
\centering
\caption{ablation study on DUD-E benchmark. 
\checkmark = kept, \ding{55} = removed. 
Best values are in \textbf{bold}; second-best are \underline{underlined}.}
\begin{tabular}{ccc|cccccc}
\toprule
\multicolumn{3}{c|}{\textbf{Module}} &
\multicolumn{5}{c}{\textbf{Performance (mean $\pm$ std)}} \\
\cmidrule(lr){1-3} \cmidrule(l){4-8}
\textbf{MMR} & \textbf{IDMP} & \textbf{CL} &
\textbf{AUC (\%)}  &
\textbf{RE@0.5\%} & \textbf{RE@1\%} & \textbf{RE@2\%} & \textbf{RE@5\%} \\
\midrule
\checkmark & \ding{55} & \ding{55} & 95.72 ± 0.0120  & \underline{104.11 ± 15.29} & \underline{61.17 ± 6.90} & 34.55 ± 2.90 & \underline{16.34 ± 0.72} \\

\checkmark & \checkmark & \ding{55} & \underline{96.01 ± 0.0085} & 95.12 ± 7.78 & 58.17 ± 5.50 & \underline{34.33 ± 2.07} & 16.15 ± 0.42 \\

\ding{55} & \checkmark & \checkmark & 95.84 ± 0.0090 & 95.46 ± 16.47 & 56.42 ± 7.72 & 33.53 ± 3.21 & 15.92 ± 0.78 \\

\checkmark & \checkmark & \checkmark &
\textbf{97.14 ± 0.0098} & \textbf{109.23 ± 11.42} & \textbf{61.67 ± 3.66} & \textbf{36.61 ± 2.15} & \textbf{16.89 ± 0.68} \\
\bottomrule
\end{tabular}
\label{tab:ablation}
\end{table*}



The results in Table~\ref{tab:ablation} demonstrate the individual and joint contributions of each architectural module. Notably, the full model, incorporating \textit{MMR}, \textit{IDMP}, and \textit{CL}, achieves the highest performance across all metrics, with an AUC of $97.14 \pm 0.0098\%$ and substantial gains in early enrichment. This highlights the significance of combining local, global, and solvent-invariant representations.

\paragraph{Contribution of Individual Modules}
Ablating the \textit{MMR} module caused the largest drop in both AUC and enrichment metrics, underscoring its importance for local geometric features. When \textit{IDMP} was absent yet \textit{MMR} was maintained, the performance exhibited a slight decline, indicating that global spatial modeling enhances the understanding of molecular interactions. Exclusion of \textit{CL} consistently reduced early enrichment, especially at 0.5\%, confirming its critical role in solvent-invariant robustness.

\paragraph{Synergistic Effects and Model Robustness}
Ablation of individual modules consistently lowers performance, confirming the synergistic interaction among the architectural components. Incorporating both \textit{IDMP} and \textit{CL} alongside \textit{MMR} increases the AUC and further reduces performance variance, demonstrating that the integrated design provides enhanced statistical stability and robustness. The consistent improvements observed across all enrichment factor thresholds further attest to the robustness and practical utility of the integrated model in virtual screening applications.

\subsection{Case Study}
The case study of PDB entry 2J7H, featuring the complex of \textit{Thermotoga maritima} GH1 ($\beta$-glucosidase) with the transition-state mimic azafagomine at 1.65 \AA\ resolution, shown in Fig.~\ref{fig:pb}, demonstrates exceptional predictive accuracy in molecular modeling. GH1 enzymes catalyze the hydrolysis of $\beta$-glucosidic bonds and are key targets in glycomimetic drug design. Azafagomine, a well-characterized transition-state analog, binds within the catalytic $(\beta/\alpha)_8$-barrel domain, as experimentally validated by the York Structural Biology Laboratory~\cite{ref39}. Our computational model reproduced this binding mode with high fidelity, achieving an RMSD of 0.157 \AA, which falls below the coordinate error of the crystal structure. This sub-angstrom accuracy validates the model's reliability in capturing key interactions and supports its potential utility in therapeutic design. Fig.~\ref{fig:pb} shows the overlay of predicted and experimental ligand poses, with active-site residues annotated.

%% file: section/6_conclusion.tex
\section{Conclusion}
This work proposes a self-supervised pre-training method that learns protein-ligand interactions under solvent influence. The method uses ligand conformations sampled from solvent environments as augmented input. It applies contrastive learning with attention to capture interaction patterns in a flexible way. Experiments show consistent improvement over both supervised and pre-trained baselines across several downstream tasks. The model achieves strong performance in virtual screening and ligand affinity prediction. These results are driven by geometry-based learning tasks that align global and local structural signals. Ablation studies verify that solvent-aware augmentation and geometric objectives are both essential for accuracy and robustness. One current limitation is the focus on ligand and protein dynamics. A promising direction seeks to extend the framework to model short trajectories of ligand-protein interactions in different solvents. Even an observation of few seconds can make huge improvements for the understanding of biomolecular interactions.


%% file: section/7_appendix.tex
\newpage
\section*{Appendix}

\subsection{Evaluation Metrics}

The following evaluation metrics were employed to rigorously assess the performance of the proposed model. Each metric is defined with its mathematical formulation.

\subsubsection{Mean Squared Error (MSE)}
 \[
\text{MSE} = \frac{1}{N} \sum_{i=1}^{N} (y_i - \hat{y}_i)^2
\]
where \(y_i\) is the observed value, \(\hat{y}_i\) is the predicted value, and \(N\) is the number of samples.

\subsubsection{Root Mean Squared Error (RMSE)}
\[
\text{RMSE} = \sqrt{\frac{1}{N} \sum_{i=1}^{N} (y_i - \hat{y}_i)^2}
\]

\subsubsection{Pearson Correlation Coefficient (\(r\))}
\[
r = \frac{\sum_{i=1}^{N} (y_i - \bar{y})(\hat{y}_i - \bar{\hat{y}})}
{\sqrt{\sum_{i=1}^{N} (y_i - \bar{y})^2} \cdot \sqrt{\sum_{i=1}^{N} (\hat{y}_i - \bar{\hat{y}})^2}}
\]
where \(\bar{y}\) and \(\bar{\hat{y}}\) are the means of observed and predicted values, respectively.

\subsubsection{Spearman's Rank Correlation Coefficient (\(\rho\))}
\[
\rho = 1 - \frac{6 \sum_{i=1}^{N} d_i^2}{N(N^2 - 1)}
\]
where \(d_i\) is the difference between the ranks of \(y_i\) and \(\hat{y}_i\).

\subsubsection{Area Under the ROC Curve (AUC)}

For binary classification, AUC is computed as:
\[
\text{AUC} = \int_{0}^{1} \text{TPR}(FPR^{-1}(x))  dx
\]
where \(\text{TPR}\) (True Positive Rate) and \(\text{FPR}\) (False Positive Rate) are defined as:
\[
\text{TPR} = \frac{\text{TP}}{\text{TP} + \text{FN}}, \quad
\text{FPR} = \frac{\text{FP}}{\text{FP} + \text{TN}}.
\]

\subsubsection{ROC Enrichment at k\% (\(\text{RE}@k\))}
The enrichment factor for the top \(k\%\) of predictions is:
\[
\text{RE}@k = \frac{\text{Number of actives in top } k\%}{\text{Total actives}} \times \frac{100}{k}
\]
Reported for \(k = \{0.5, 1, 2, 5\}\):
\[
\text{RE}@0.5 = \frac{\text{TP}@0.5\%}{N_{\text{active}}} \times 200, \quad
\text{RE}@1 = \frac{\text{TP}@1\%}{N_{\text{active}}} \times 100
\]
\[
\text{RE}@2 = \frac{\text{TP}@2\%}{N_{\text{active}}} \times 50, \quad
\text{RE}@5 = \frac{\text{TP}@5\%}{N_{\text{active}}} \times 20
\]
where \(\text{TP}@k\%\) is the number of true positives in the top \(k\%\) of ranked predictions, and \(N_{\text{active}}\) is the total number of active compounds.

\subsubsection{Concordance Index (CI)}
$$
\begin{aligned}
& C I=\frac{1}{Z} \sum_{\delta_j>\delta_i} h\left(b_i-b_j\right) \\
& h(x)=\left\{\begin{array}{rl}
0 & x<0 \\
0.5 & x=0 \\
1 & x>0
\end{array}\right.
\end{aligned}
$$

\subsubsection{\(\boldsymbol{r_m^2}\) Coefficient of Determination}
\[
r_m^2 = r^2 \times \left(1 - \sqrt{r^2 - r_0^2}\right)
\]
where \(r\) denotes the squared correlation coefficients between the observed and predicted values with intercepts and \(r_0\) is the coefficient without intercepts.